\documentstyle[12pt]{article} 
\begin{document}
 \begin{center}
 {\bf  Energy and time as conjugate dynamical variables} \\[2cm] 
Marius Grigorescu \\[3cm]  
\end{center}
 {\small Abstract}: The energy and time variables of the elementary classical dynamical systems are described geometrically, as canonically conjugate coordinates of an extended phase-space. It is shown that the Galilei action of the inertial equivalence group on this space is canonical, but not Hamiltonian equivariant. Although it has no effect at classical level, the lack of equivariance makes the Galilei action inconsistent with the canonical quantization. A Hamiltonian equivariant action can be obtained by assuming that the inertial parameter in the extended phase-space is quasi-isotropic. This condition leads naturally to the Lorentz transformations between moving frames as a particular case of symplectic transformations. The limit speed appears as a constant factor relating the two additional canonical coordinates to the energy and time. Its value is identified with the speed of light by using the relationship between the electromagnetic potentials and the symplectic form of the extended phase-space. \\[1cm] 
\newpage
 {\bf 1. Introduction} \\[.5cm] \indent
 The canonical transformations of the Hamilton-Jacobi theory have the remarkable property of changing the energy as a variable canonically conjugate with time. This formal property is consistent with generalized Hamiltonian dynamics in an extended phase-space, where the energy and time are true dynamical variables, rather then parameters \cite{macke}. The canonical transformations in the extended phase-space provide the appropriate framework to describe the equivalence of inertial frames in relative motion, and include the Lorentz transformations as a special case \cite{macke}. \\ \indent
 A clear distinction between time and the other phase-space variables appears in the quantum theory.  The time remains classical, while the energy, the momenta, and the space coordinates are affected by statistical fluctuations. However, the uncertainty relations between proper time and mass suggest that time and energy should be also considered as canonically conjugate dynamical variables \cite{km}. \\ \indent 
The phase-space structure of the classical dynamical systems can be described geometrically, by using an associated symplectic manifold $(M, \omega)$. Here $M$ denotes the manifold of the physical states,
while $\omega$ is the symplectic form on $M$. For conservative systems, the dynamics is a symplectic diffeomorphism of $(M, \omega)$, generated by the Hamilton function $H$. The existence of a generating function ensures that this diffeomorphism can be described globally by only one parameter, but this parameter, (the time), does not appear explicitly. Therefore, a geometrical formulation of the Hamilton-Jacobi theory requires the extension of the symplectic manifold $M$ to the contact manifold $M \times R$, where $R$, as additive group, denotes the time axis \cite{am}. The contact structure is derived from the symplectic structure of $M$, by considering the Hamiltonian as conjugated to the time coordinate.
However, in this approach $H$ does not appear as an additional coordinate, but the one on $M$ defined by foliation with constant energy surfaces. \\ \indent 
In this work the classical dynamics of elementary systems is formulated in terms of an extended phase-space $M \times T^*R$, where $T^*R$ denotes the cotangent bundle of the time axis. The time-dependent Hamilton-Jacobi theory, and examples related to the transformations between moving frames will be presented in Sect. 2. \\ \indent 
The symplectic structure of the extended phase-space provides the appropriate framework for the treatment of the coupling between a charged particle and an electromagnetic field. In the early attempts to formulate a geometrical theory of electromagnetism, the 4-vector potential was supposed to be related to the metric structure of the space-time \cite{weyl}. However,  later developments strongly suggest that electromagnetism is naturally related to the symplectic structure of phase-space ( \cite{gs} p. 140). In Sect. 3 it will be shown that the coupling between a charged particle and the electromagnetic field can be described in terms of a potential-dependent symplectic form on $M \times T^*R$. The Lorentz force and the homogeneous Maxwell equations follow as consequences. \\ \indent 
The relationship between inertial frames in relative motion (the inertial equivalence group) is studied in Sect. 4. It will be shown that the Galilei action in extended phase-space is canonical, but not Hamiltonian equivariant. This means that it is not possible to find a homomorphism from the Lie algebra of the Galilei group to the Poisson algebra ${\cal F} (M)$ of the smooth real functions on $M$. This homomorphism does not exist if the transition to a moving frame (boost) acts by shifting the momenta with a velocity-dependent term. However, the obstruction disappears if the action of the inertial equivalence group is supposed to preserve the canonical structure of the extended phase-space, and if the inertial parameter is quasi-isotropic. With these assumptions, the velocity-dependent momentum shift is replaced by a symplectic transformation, and the Galilei action by the Poincar\'e transformations. Therefore, to obtain the Lorentz transformations, beside the conditions presented in ref. \cite{macke}, it is necessary to assume the quasi-isotropy of the inertial parameter in the extended phase-space. The main results and conclusions are summarized in Sect. 5. \\[.5cm] 
{\bf 2. Time-dependent canonical transformations} \\[.5cm] \indent 
In Hamilton-Jacobi theory, a local representation of the time-dependent symplectic diffeomorphisms that transform a Hamiltonian vector field into another Hamiltonian vector field is provided by the generating function of the canonical transformations. Such diffeomorphisms may be pictured as transformations, generated by the action field $S(q,p',t)$, of a dynamical system with coordinates $(q,p)$, and Hamilton function $H$, to a "moving frame", with coordinates $(q',p')$. For a system with $n$ degrees of freedom the action field can be expressed in the form 
\begin{equation} S( \tilde{q}, \tilde{p}',t) = \tilde{q}^T \tilde{p}' + \Phi ( \tilde{q}, \tilde{p}',t)~~, 
\end{equation} 
where $\tilde{q}$ and $\tilde{p}'$ denote the column vectors with 
components $q^k$, $p'_k$, respectively, $k=1,n$. This leads to 
an implicit relationship between coordinates, 
\begin{equation} 
p_k = p'_k + \frac{ \partial \Phi}{ \partial q^k} ~~,~~ q'^k = q^k + \frac{ \partial \Phi }{ \partial p'_k}
\end{equation} 
which preserve the structure of the equations of motion, such that the time evolution of the new coordinates is generated by the Hamilton function 
\begin{equation} 
H'= H + \frac{ \partial \Phi }{ \partial t}~~. 
\end{equation} 
The identity transformation is generated by the functions $\Phi$, which are constants. For infinitesimal transformations close to identity, $\Phi$ has the general form 
\begin{equation} 
\Phi ( \tilde{q}, \tilde{p}',t) = \tilde{X}^T \tilde{q} - \tilde{Y}^T \tilde{p}' + \frac{1}{2} (\tilde{q}^T \hat{b} \tilde{q} + \tilde{p}'^T \hat{c} \tilde{p}' ) - \tilde{q}^T \hat{a} \tilde{p}' 
\end{equation} 
where $\tilde{X}$, $\tilde{Y}$ are $n$-component column vectors, and $\hat{a}$, $\hat{b}=\hat{b}^T$, $\hat{c}=\hat{c}^T$ are $n \times n$ matrices. The transformation defined by (2) takes, in this case, the matrix form 
\begin{equation} 
\left[ \begin{array}{c} \tilde{q}' \\ \tilde{p}' \end{array} \right] = \left[ \begin{array}{c} \tilde{q} \\ \tilde{p} \end{array} \right] + \left[ \begin{array}{c} - \tilde{Y} \\ - \tilde{X} \end{array} \right] + \left[ \begin{array}{cc} - \hat{a}^T & \hat{c} \\ - \hat{b} & \hat{a} \end{array} \right] 
\left[ \begin{array}{c} \tilde{q} \\ \tilde{p} 
\end{array} \right]~~. 
\end{equation} 
Therefore, $\Phi$ generates a momentum shift parameterized by $\tilde{X}$, a coordinate shift parameterized by $\tilde{Y}$, and a symplectic linear transformation generated by an element of $sp(n,R)$, parameterized by $\hat{a}$, $\hat{b}$, and $\hat{c}$. \\  \indent 
The infinitesimal transformations between frames in relative motion are particular cases of this general formula. The dynamics of an $N$-particle many-body system in a rotating frame can be expressed in terms of the new canonical variables $\tilde{q'}=[{\bf q'}_i, i=1,N]$, $\tilde{p'}= [ {\bf p'}_i, i=1,N ] $, defined by (5), with $\tilde{X} = \tilde{Y} = \tilde{0}$, $\hat{b}=\hat{c}=0$ and $\hat{a} = \otimes_{i=1}^N ( \xi )_i$, $\xi \in so(3)$. If $\Omega_\mu$, $\mu=1,2,3$, denotes the components of the angular velocity vector, and $\epsilon_{ \sigma \mu \nu}$ is the Levi-Civita symbol, then $\xi_{ \mu \nu} = \delta t \sum_{ \sigma =1}^3 \Omega_{ \sigma} \epsilon_{ \sigma \mu \nu}$. The generating function of this transformation is $\Phi_r ( \tilde{q}, \tilde{p}', \delta t) = - \delta t {\bf \Omega} \cdot {\bf L}$, where ${\bf L}= \sum_{i=1}^N {\bf q}_i \times {\bf p'}_i$. The new coordinates are related to the initial ones by 
\begin{equation} 
{\bf q}' = {\bf q} - \delta t {\bf  \Omega} \times {\bf q} ~~,~~  {\bf p}' = {\bf p} - \delta t {\bf  \Omega} \times {\bf p} \end{equation} 
and their time evolution is generated by the Hamiltonian $H'=H- {\bf \Omega} \cdot {\bf L}$. \\ 
\indent The transformation to a frame in uniform translation with the relative velocity ${\bf V}$ corresponds to $\hat{a}=\hat{b}=\hat{c}=0$ and $\tilde{X} = \sum_{i=1}^N {\bf X}_i$, $\tilde{Y} = \sum_{i=1}^N {\bf Y}_i$, with ${\bf X}_i = m_i {\bf V}$, ${\bf Y}_i = t {\bf V} $. In this case the generating function is 
\begin{equation} 
\Phi_v ( \tilde{q}, \tilde{ p}',t)= {\bf V} \cdot (m {\bf q}_{cm} - t {\bf P}' ) 
\end{equation} 
where $m$ is the total mass of the system, ${\bf q}_{cm}$ is the center of mass coordinate in the old frame and ${\bf P}'$ is the total momentum in the new frame. The Hamilton function in the new frame is 
\begin{equation} 
H'( \tilde{q}', \tilde{p}',t) = H({\bf q}'_i+ t {\bf V}, {\bf p}'_i+m_i {\bf V}, t) - {\bf V} \cdot {\bf P}' ~~.
\end{equation} 
\indent The Hamiltonian dynamics of an elementary system with $n$ degrees of freedom defined on $(M, \omega)$ can be seen as the action of the additive group $R$, parameterized by time. Considering the Hamiltonian $H \in \cal{F} (M)$ as the momentum mapping for this action, the energy $E \in R$ appears as an element of the dual of the "time-group" algebra, $R$. The time and energy variables $(t, E)$ therefore can be treated as coordinates on the cotangent space $T^*R$. The canonical coordinates on $T^*R$, denoted by $(q^0,p_0)$, are supposed to be linear functions of energy and time, $q^0= c t$, $p_0 = - E /c$, with $c$ a dimensional constant. With these new canonical variables, the extended phase-space is represented by the symplectic manifold $(M^e, \omega^e_0)$, $M^e = M \times T^* R$ and $\omega^e_0 = \omega + dq^0 \wedge dp_0$. \\ \indent 
The Hamiltonian current $X_H$ on $(M, \omega)$, generated by $H$, can be lifted to a Hamiltonian current $X^e_H$ on $(M^e, \omega^e_0)$, generated by $H^e= H+cp_0 \in \cal{F} (M^e)$. Let $s$ be the parameter chosen along the trajectories on $M^e$, $d_s \equiv d /d s $ the derivative with respect to $s$, $L_{X^e_H} = \sum_{i=1}^n (d_s q)^i \partial_{q^i} + (d_s p)_i \partial_{p_i} + (d_s t ) \partial_{ t } + (d_s E) \partial_{E}$ the Lie derivative associated to $X^e_H$, $\omega= \sum_{i=1}^n dq^i \wedge dp_i$, and $i_{X^e_H} \omega^e_0$ the inner product between $X^e_H$ and $\omega_0^e$. Thus, if $X^e_H$ is the current generated by the Hamiltonian $H^e$, then $i_{X^e_H} \omega^e_0 = d H^e$, and the corresponding equations of motion in the extended phase-space have the form
\begin{equation} d_s q^i = \frac{ \partial H}{ \partial p_i} ~~,~~ d_s p_i = - \frac{ \partial H}{ \partial q^i} 
\end{equation} 
\begin{equation} d_s t = 1 ~~,~~ d_s E = \frac{ \partial H}{ \partial t } 
\end{equation} 
The first group of equations are the usual Hamilton equations on $(M, \omega)$. The second group shows that the choice of $H^e$ corresponds to $s =t$, and ensures the conservation of the energy when $H$ is independent of $t$. \\ \indent 
The generating functions of the canonical transformations in the extended phase-space are supposed to be independent of $s$. It is easy to check that the equations of motion of an elementary system in uniformly rotating or translating frames, can be derived from Eq. (9), (10), by a canonical transformation generated by $\Phi_r$ or $\Phi_v$, respectively. 
\\[.5cm] 
{\bf 3. The coupling to the electromagnetic field} \\[.5cm] 
\indent 
The natural symplectic form on the one-particle phase-space $M=T^*R^{3}$ can be modified to describe the coupling to the electromagnetic field \cite{gs}. However, a suitable framework to state this procedure is represented by the extended phase-space, $M^e= T^* R^4$. In a Coulomb field, the energy $E$ of a charged particle becomes $E+e V$, where $e$ is the charge and $V$ the potential. This corresponds to a change of $p_0$ to $p_0- eV/c$, where $V$ can be a general function of coordinates and time. The prescription of introducing the field by a local shift of the $p_0$ axis can be generalized to all momentum components of $\omega^e_0$. Thus, in the presence of the electromagnetic field, $\omega^e_0$ is replaced by 
\begin{equation} 
\omega^e({\bf A}, V) = \sum_{\mu =1}^3 d {q}_\mu \wedge d( {p}_\mu + \frac{e }{c} {A}_\mu {({\bf q},t)} ) + dq^0 \wedge d( p_0 - \frac{e}{c} V {( {\bf q},t)} ) 
\end{equation} 
where ${\bf A}$ is a function of the coordinates and time representing the vector potential. This procedure is naturally gauge-invariant, because the form $\omega^e( {\bf A}, V)$ remains the same at the transformation 
\begin{equation} 
{\bf A} \rightarrow {\bf A} + {\bf  \bigtriangledown} f ~~,~~ V \rightarrow V - \frac{1}{c} \frac{ \partial f}{ \partial t} \end{equation} 
where $f$ is an arbitrary function of coordinates and time. \\ \indent The field contribution to the symplectic form of $M^e$ can be separated in magnetic and electric components, given by the decomposition $ \omega^e({\bf A}, V) = \omega^e_0 + \omega_B + \omega_E$, where 
\begin{equation} 
\omega_B=- \frac{e}{c}(B_1 dq_2 \wedge dq_3 +B_2dq_3 \wedge dq_1+B_3 dq_1 \wedge dq_2 ) 
\end{equation} 
\begin{equation} 
\omega_E = - e {\bf E} \cdot d {\bf q} \wedge dt 
\end{equation} 
and 
\begin{equation} 
{\bf B} = {\bf  \bigtriangledown } \times {\bf A}~~,~~ {\bf E} = - \frac{1}{c} \frac{ \partial {\bf A}}{ \partial t}- {\bf  \bigtriangledown } V ~~,~~ {\bf  \bigtriangledown } \equiv 
{\bf  \bigtriangledown_q} = \frac{ \partial}{ \partial {\bf q}}~.
\end{equation} 
If $X_H^e$ denotes the Hamiltonian current in the extended phase-space, then 
\begin{equation} 
i_{X_H^e} \omega^e_0 = \sum_{\mu =1}^3 ( \dot{{q}}_\mu d {p}_\mu - \dot{{p}}_\mu d {q}_\mu ) - d E + \dot{ E} dt 
\end{equation} 
\begin{equation} 
i_{X_H^e} \omega_B = \frac{ e}{c} (\dot{ {\bf q} } \times {\bf B} ) \cdot d {\bf q} 
\end{equation} 
\begin{equation} 
i_{X_H^e} \omega_E = - e ( {\bf E} \cdot \dot{{\bf q}} dt - {\bf E} \cdot d {\bf q} ) 
\end{equation} 
such that the equations of motion $i_{X_H^e} \omega^e( {\bf A},V) = dH^e$ are 
\begin{equation} 
\dot{{\bf q} } = {\bf  \bigtriangledown_p} H~~,~~ \dot{ {\bf p}} = - {\bf  \bigtriangledown_q} H + e [ \frac{1}{c} \dot{ {\bf q}} \times {\bf B} + {\bf E} ] 
\end{equation} 
and 
\begin{equation} 
\frac{ d E}{dt} = e \dot{ {\bf q}} \cdot {\bf E} + \frac{ \partial H}{ \partial t} ~~.
\end{equation} 
The first two equations are the classical equations of motion containing the velocity-dependent Lorentz force term, while the last equation expresses the rate of change of the mechanical 
energy of the particle. The two homogeneous Maxwell equations 
\begin{equation} 
{\bf  \bigtriangledown} \cdot {\bf B} = 0 ~~,~~ {\bf  \bigtriangledown} \times {\bf E} = - \frac{ \partial {\bf B}}{ \partial t}
\end{equation} 
are a consequence of the definition (15). The gauge invariance can be used to fix ${\bf A}$ and $V$ such that $c {\bf  \bigtriangledown} \cdot {\bf A} + \partial V / \partial t =0$, and in this case equations (15) lead to 
\begin{equation} 
{\bf  \bigtriangledown} \times {\bf B} - \frac{1}{c} \frac{ \partial {\bf E}}{ \partial t} = \frac{1}{ c^2} \frac{ \partial^2 {\bf A}}{ \partial^2 t} - \bigtriangleup {\bf A} ~~.
\end{equation} 
If the homogeneous Amp\`ere-Maxwell equation 
\begin{equation} 
{\bf  \bigtriangledown} \times {\bf B} - \frac{1}{c} \frac{ \partial {\bf E}}{ \partial t} =0 
\end{equation} 
is taken as the definition of the vacuum, then (22) shows that ${\bf A}$ satisfies the wave equation, and the constant $c$ is the speed of light in vacuum. \\ \indent 
It is interesting to remark that by introducing new momentum coordinates $( {\bf  p}', p_0') \equiv ( {\bf p} + e {\bf A} /c, p_0-e V/c)$, the symplectic form $\omega^e( {\bf A}, V)$ takes the form of $\omega^e_0$, while the Hamiltonian $H^e= H({\bf q},{\bf p},t)-E$ becomes $H^e= H({\bf q}, {\bf p'} - e {\bf A} /c,t)-E'+ e V $. Therefore, in the new coordinates the dynamics is Hamiltonian, and the field-dependent Poisson bracket in the extended phase-space, $\{ *,* \}^e_f$, defined with respect to $(q^0, {\bf q})$ and $(p_0', {\bf p'})$, is independent of time. Moreover, the old coordinates and momenta satisfy the relations $\{q_\mu,p_{ \nu} \}^e_f= \delta_{ \mu \nu}$, $\{q_\mu,q_{ \nu} \}^e_f= 0$, $\{p_\mu,p_{ \nu} \}^e_f = - e \epsilon_{ \mu \nu \sigma} B_{ \sigma} /c$, $\{p_\mu,p_0 \}^e_f = e E_{ \mu} /c$. These properties can be used to formulate a classical version of the Feynman proof of the homogeneous Maxwell equations \cite{fd}. The commutator from the Feynman's proof corresponds in the classical case to the field-dependent, invariant Poisson bracket $\{*,* \}^e_f$, instead of $\{ *, * \}^e$ defined in terms of $(q^0, {\bf q})$ and $(p_0, {\bf p})$. Thus, it is possible to show that the most general velocity-dependent force compatible with the Newton's equations of motion, and ensuring the existence of an invariant Poisson bracket $\{*,* \}^e_f$, such that $\{q_\mu,p_\nu \}^e_f= \delta_{ \mu \nu}$ and $\{q_\mu, q_\nu \}^e_f= 0$, is the Lorentz force. 
\\[.5cm] 
{\bf 4. Symplectic actions of the inertial equivalence group} \\[.5cm] 
\indent 
The momentum variables are changed not only by the coupling to the electromagnetic field, but also by the transition to a moving frame. The transformation to a coordinate frame in uniform motion generated by the function $\Phi_v$ of (7) is a special case of the general Galilei transformation $\Gamma_Q : R^3 \times R \rightarrow R^3 \times R$ acting both on the coordinate space $R^3$ and time. If $({\bf q}, {\bf p})$ denote the Cartesian phase-space coordinates of a particle with mass $m$, then an infinitesimal Galilei transformation is defined by $ [ {\bf q'}, t']= [ {\bf q}, t ]+ \gamma (\xi, {\bf d}, {\bf v}, \tau) [ {\bf q},t ]$, with 
\begin{equation} 
\gamma (\xi, {\bf d}, {\bf v}, \tau) [ {\bf q} , t ] = [\xi {\bf q} - {\bf d} - t {\bf v}, - \tau]~~. 
\end{equation} 
\indent The algebra $g$ of the Galilei group is isomorphic to $so(3) + R^7$. An element $\gamma \in g$ is specified by $ \xi \in so(3)$, ${\bf d} \in R^3$, ${\bf v} \in R^3$ and $\tau \in R$. The parameters $\xi$, ${\bf d}$ and ${\bf v}$ correspond to static rotations, translations and boost, respectively, of the space coordinates, while $\tau$ describes translations along the time axis. \\ \indent 
The action $\Gamma_Q$ of the Galilei group can be lifted to an action $\Gamma_M$ on the phase-space $M=T^* R^3$, by assuming that at the transformation specified by (24), the momentum also changes as ${\bf p}'= {\bf p} + \xi {\bf p} - m {\bf v}$. Let ${\cal D}(M)$ be the group of diffeomorphisms of $M$, and ${\cal D}(M, \omega) = \{ \rho \in {\cal D}(M), \rho^* \omega = \omega \}$ the subgroup of symplectic diffeomorphisms. When $M= T^* R^{3}$ and $\omega = \sum_{\mu=1}^3 d {q}_\mu \wedge d {p}_\mu$, then $\Gamma^*_M \omega = \omega$, and the action $\Gamma_M$ of the Galilei group is symplectic. \\ 
\indent 
By the action $\Gamma_M$ each element $\gamma \in g$ generates a current $X_\gamma \in TM$. The Lie derivative associated with the current $X_v$ generated by the boost transformation $\gamma_v \equiv \gamma (0, 0, {\bf v}, 0) \in g$ is 
\begin{equation} L_{X_v} = - {\bf v} \cdot (t {\bf  \bigtriangledown_q} + m  {\bf  \bigtriangledown_p}) ~~. 
\end{equation} 
This current is Hamiltonian, and satisfies the equation $i_{X_v} \omega = d J_v (q,p,t)$, where $J_v (q,p,t) = \Phi_v (q,p,t)= {\bf v} \cdot (m {\bf q}-t {\bf p})$. The current $X_d $, corresponding to the static shift of the origin $\gamma_d \equiv \gamma (0, {\bf d}, 0, 0) \in g$ has the associated Lie derivative $ L_{X_d} = - {\bf d} \cdot {\bf  \bigtriangledown_q} $, and is generated by $J_d = - {\bf d} \cdot {\bf p}$. Similarly, a Hamilton function $J_\gamma$ exists for every vector field $X_\gamma$, $\gamma \in g$, and therefore, the action $\Gamma_M$ induces an anti-homomorphism $d \Gamma_M: g \rightarrow ham(M)$, $d \Gamma_M ( \gamma) = X_\gamma$, $X_{[ \gamma, \gamma']} = - [ X_\gamma, X_{\gamma'}]$, between the Lie algebra $g$ of the Galilei group and the Lie algebra $ham(M)$ of the Hamiltonian vector fields on $M$ (ref. \cite{am} p. 269). \\ \indent 
The action $\Gamma_M$ will be called Hamiltonian equivariant, if the Lie
algebra anti-homomorphism $d \Gamma_M : g \rightarrow ham(M)$ can be lifted to a homomorphism $\lambda : g \rightarrow {\cal F} (M)$, such that the diagram 
\\ 
\begin{center} 
\( \begin{array}{ccc}
 0 ~ \rightarrow ~R ~  \rightarrow & {\cal F} (M) ~ \rightarrow  ~ ham(M)  &  \rightarrow ~ 0 \\ 
~~~~~~~ & ~~~~~_{\lambda}  \nwarrow ~~  \uparrow_{d \Gamma_M} & \\ 
~~~~~~~ & ~~~~~~~~ g &  
\end{array} \) 
\\ 
\end{center} 
commutes. This property can be expressed in a compact form by the equation $[{\cal Q}]=0$, where $[{\cal Q}] \in H^2(g,R) $ is the cohomology class of the two-cocycle defined by ${\cal Q}( \gamma ; \gamma ') = \{ J_{\gamma}, J_{\gamma '} \} - J_{ [ \gamma , \gamma ' ] }$, $\gamma, \gamma ' \in g$  (ref. \cite{gs} p. 171). \\ \indent
The equivariance is necessary for a consistent quantization, because the physical observables are elements of ${\cal F} (M)$, rather than of $ham (M)$. Therefore, it is possible to find a representation of $g$ by operators associated with the observables of the particle, acting on the quantum Hilbert space $L^2(R^3)$, only if $\lambda$ exists. \\ \indent  
In the case of the lift $\Gamma_M$, the commutator $[ \gamma_v , \gamma_d ] $ is 0, but the Poisson bracket of $J_v$ and $J_d$ is 
\begin{equation} 
\{ J_v, J_d \} = - L_{X_v} J_d = - m \{ {\bf v} \cdot {\bf q} ,{\bf d} \cdot {\bf p} \} = - m {\bf v} \cdot {\bf d}~~. 
\end{equation} 
Therefore, although the infinitesimal Galilei transformations $\gamma(0, {\bf d}, {\bf v}, 0)$ and $\gamma'(0, {\bf d}', {\bf v}', 0)$ commute, the Poisson bracket of the corresponding Hamilton functions 
\begin{equation} 
\{ J_{\gamma}, J_{\gamma'} \} = m ( {\bf d} \cdot {\bf v}' - {\bf d}' \cdot {\bf v}) 
\end{equation} 
is, in general, not zero. This Poisson bracket defines a two-cocycle ${\cal Q}$ on $g$, ${\cal Q}( {\bf v}, {\bf d} ; {\bf v}', {\bf d}') = \{ J_{\gamma}, J_{\gamma} ' \}$, parameterized by the mass $m$ (refs. \cite{jms} and \cite{gs}, p. 434). Thus, the cohomology class $[{\cal Q}]$ of ${\cal Q}$ in $H^2(g,R)$ is zero, and the action of the inertial equivalence group is Hamiltonian equivariant, only if $m=0$. \\ \indent 
When $m \ne 0$, the lack of equivariance and the obstruction in finding $\lambda$ are due to the noncommutation of the coordinates and momenta with respect to the Poisson bracket. The diagram commutes if $\lambda$ maps the generators of the phase-space translations onto the phase-space coordinates. However, this mapping is not a homomorphism, because the Lie algebra of the translation group is Abelian, while the phase-space coordinates $q_\mu, p_\nu$, with respect to the Poisson bracket, generate the Heisenberg algebra, $\{ q_\mu, q_\nu \}=0$, $\{ p_\mu, p_\nu \}=0$, $\{ q_\mu, p_\nu \}= \delta_{ \mu \nu}$, $\mu, \nu =1,2,3$. 
\\ \indent
The boost transformations depend on time explicitly, and therefore the action of the Galilei group can be formulated in terms of the extended phase-space. If the coordinates on $M^e$ are represented as column vectors $$ \tilde{q} = \left[ \begin{array}{c} {\bf q} \\ q^0 \end{array} \right]~~,~~ \tilde{p} = \left[ \begin{array}{c} {\bf p} \\ p_0 \end{array} \right] $$ 
then an infinitesimal transformation $\Gamma_M$ takes the form of (5), with 
$$ \tilde{X} = \left[ \begin{array}{c} m {\bf v} \\ 0 \end{array} \right] ~,~ \tilde{Y} = \left[ \begin{array}{c} {\bf d} \\ \tau \end{array} \right] ~,~ \hat{a} = \left[ \begin{array}{cc} \xi & {\bf 0} \\ {\bf v}/c & 0 \end{array} \right]~, $$
and $\hat{b}=\hat{c}=0$. The element ${\bf v}/c$ of the matrix $\hat{a}$ is determined only by $\Gamma_Q$, but according to (5), its presence requires the transformation of $p_0=-E/c$ as $p_0'=p_0 + {\bf v} \cdot {\bf p} /c $. This transformation is consistent with (8), and it leads to the correct equations of motion in the new frame. Therefore, the action $\Gamma_Q$ can be lifted to a symplectic action $\Gamma_{M^e}$ in the extended phase-space. \\ \indent 
The Lie derivative associated to the current $X^e_v= d \Gamma_{M^e} ( \gamma_v )$ is 
\begin{equation} 
L_{X^e_v} = - q^0 \frac{ {\bf v}}{c} \cdot {\bf  \bigtriangledown_q} - m {\bf v} \cdot {\bf  \bigtriangledown_p} + \frac{ {\bf p} \cdot {\bf v}}{c} \frac{ \partial}{ \partial p_0} ~~, 
\end{equation} 
and $i_{X^e_v} \omega^e_0 = dJ^e_{v}$, with $J^e_v = {\bf v} \cdot (m {\bf q}-q^0 {\bf p}/c)$. The current $X^e_d$ determined by $\gamma_d $ has the generating function $J^e_d = - {\bf d} \cdot {\bf p}$, and similarly to (26), 
\begin{equation} 
\{ J^e_v, J^e_d \}^e = - m {\bf v} \cdot {\bf d} ~~. 
\end{equation} 
Therefore, the action $\Gamma_{M^e}$ of the Galilei group in the extended phase space defines the same element [${\cal Q}$] of $H^2(g,R)$ as $\Gamma_M$, and is not Hamiltonian equivariant. \\ \indent 
The noncommutation of the phase-space coordinates affects the equivariance because $\gamma_v$ acts by shifting the momentum components. This shift ($ \tilde{X}$) is proportional to the inertial parameter $m$, defined by the momentum dependence of the Hamiltonian. In principle, each momentum component $p_k$ has an associated inertial parameter $ m_k = p_k ( \partial H^e/ \partial p_k)^{-1}$. If the dynamics in the extended phase-space is determined by $H^e$, then it is natural to assume that the corresponding inertial parameter is quasi-isotropic, namely $ m_1=m_2=m_3 = - \alpha m_0 =m >0$, $\alpha = \pm 1$. This assumption is equivalent to a relationship between mass and energy of the form $ E= \alpha m c^2$. Therefore, $m= - \alpha p_0/c$, and the lift ${\bf p'}= {\bf p} + \xi {\bf p} - m {\bf v}$ of $\gamma_v$ places the velocity-dependent term in the matrix $\hat{a}$, instead of $\tilde{X}$. The infinitesimal transformation to a moving frame takes in this case the form of (5), with 
$$ \tilde{X} = \left[ \begin{array}{c} 0 \\ 0 \end{array} \right] ~~,~~ \tilde{Y} = \left[ \begin{array}{c} {\bf d} \\ \tau \end{array} \right] $$ and 
\begin{equation} 
\hat{a} = \left[ \begin{array}{cc} \xi & \alpha {\bf v} /c \\ {\bf v}/c & 0 \end{array} \right]~~. 
\end{equation} 
The presence of the $\alpha$ - dependent term in the matrix $\hat{a}$ is not consistent with the action $\Gamma_Q$ defined by (24). Thus, (5) requires the transformation of the coordinates by the matrix $- \hat{a}^T$, and an action of the inertial equivalence group of the form 
$ \gamma^e (\xi, {\bf d}, {\bf v}, \tau) [ {\bf q} , t ] = [\xi {\bf q} - {\bf d} - t {\bf v}, - \alpha {\bf v} \cdot {\bf q} /c^2 - \tau] $.
For this action 
\begin{equation} L_{X^e_v} = - q^0 \frac{ {\bf v}}{c} \cdot {\bf  \bigtriangledown_q} + \alpha p_0 \frac{ {\bf v}}{c} \cdot {\bf  \bigtriangledown_p} + \frac{ {\bf p} \cdot {\bf v}}{c} \frac{ \partial}{ \partial p_0} - \alpha \frac{ {\bf v}}{c} \cdot {\bf q} \frac{ \partial}{ \partial q^0}~, 
\end{equation} 
$J^e_v= - {\bf v} \cdot ( \alpha p_0 {\bf q} + q^0 {\bf p})/c$, and 
\begin{equation} 
[ L_{X^e_v}, L_{X^e_d} ] =- \alpha \frac{ {\bf v}}{c} \cdot {\bf d} \frac{ \partial }{ \partial q^0} ~~. 
\end{equation} 
This shows that $J^e_{ [ X^e_v, X^e_d ]} =- \alpha p_0 {\bf v} \cdot {\bf d} /c$, and as $[\gamma_v^e, \gamma_d^e] = \gamma^e(0,0,0, \alpha {\bf v} \cdot {\bf d} /c ) $, 
\begin{equation} 
\{ J^e_v, J^e_d \}^e-J^e_{[ \gamma_v^e, \gamma_d^e]} = 0 ~~. 
\end{equation} 
Therefore, the cohomology class of the two-cocycle defined by Eq. (33) is zero, proving that the new action is Hamiltonian equivariant. \\ \indent 
The choice $\alpha = 1 $ corresponds to the usual relation between mass and energy, $ E = mc^2$. For uniform translations with a finite velocity ${\bf V} = V {\bf n}$, $\vert {\bf n} \vert=1$, this leads to the standard Lorentz transformations 
\begin{equation} 
{\bf q}' = {\bf n} \times ({\bf q} \times {\bf n})  + {\bf n} \frac{ {\bf q} \cdot {\bf n} -  V t}{ \sqrt{1 - V^2/c^2}  }~~,~~ t' = \frac{ t - {\bf V} \cdot {\bf  q}/c^2 }{ \sqrt{1 - V^2/c^2}} ~~. \end{equation} 
The choice $\alpha = -1 $ corresponds to $E = - mc^2$, and 
\begin{equation} 
{\bf q}' = {\bf n} \times ({\bf q} \times {\bf n})  + {\bf n} \frac{ {\bf q} \cdot {\bf n} - V t}{ \sqrt{1 + V^2/c^2}} ~~,~~ t' = \frac{ t + {\bf V} \cdot {\bf  q}/c^2 }{ \sqrt{1 + V^2/c^2}} ~~. \end{equation} 
These transformations represent pure rotations between the space coordinates and time. By contrast with the Lorentz transformations, they do not leave the vacuum defined by (23) invariant. \\ \indent
The infinitesimal transformations of the momentum components ${\bf p}'={\bf p} + \xi {\bf p} + \alpha p_0 {\bf v}/c$ and $p_0'=p_0+{\bf v} \cdot {\bf p}/c$, indicate that the inertial parameter $m= - \alpha p_0/c$ remains invariant only if ${\bf v}=0$. However, the quantity ${\bf p}^2- \alpha p_0^2$ is a general invariant, which can be used to specify the relationship between $E$ and ${\bf p}$ in any particular frame\footnote{$E= \alpha c \sqrt{m^2 c^2 + \alpha {\bf p}^2}= \alpha mc^2 / \sqrt{1- \alpha V^2/c^2}$ when ${\bf p}=m {\bf V} / \sqrt{1- \alpha V^2/c^2}$.}  
\\[.5cm] 
{\bf 5. Summary and Conclusions} \\[.5cm] 
\indent 
The invariance of the Hamiltonian dynamical systems at static canonical transformations can be well described in terms of the symplectic geometry of the phase-space. However, the time-dependent transformations require a more general formalism, which is provided by the Hamilton-Jacobi theory. In this formalism, the time appears as a dynamical variable, rather than as a parameter similar to the ones describing the static transformations. Moreover, this variable has the properties of a coordinate, having the energy as conjugate momentum. Therefore, the Hamilton-Jacobi theory suggests the extension of the phase-space by two new canonically conjugate variables $(q^0,p_0)$, related to energy and time by a dimensional factor $c$ \cite{macke}. \\ \indent 
In this work it was shown that the symplectic structure of the extended phase-space of a charged particle can be modified to take into account the coupling to the electromagnetic field. The field potentials are introduced by assuming that all momentum components are changed in the same way as $p_0$ (the energy), by an additive coordinate-dependent term. This procedure is gauge-invariant, leads to the correct expression of the Lorentz force, of the rate of change of the mechanical emergy, and it allows us to identify the constant $c$ with the speed of light. \\ \indent 
An important class of time-dependent canonical transformations is provided by the equivalence principle of the inertial frames. The group of transformations between inertial frames in relative motion (the inertial equivalence group), changes not only the coordinates, but also the momenta and energy, and therefore acts on the extended phase-space. \\ \indent 
In Sect. 4 it was shown that the Galilei transformations preserve the canonical structure of the extended phase-space. However, this action defines a two-cocycle in $H^2(g,R)$ parameterized by the mass, and is not Hamiltonian equivariant. The lack of equivariance has no effect at  classical level, while in quantum theory it affects the wave functions by a global phase factor only. Though, it shows that the Galilei action is basically inconsistent with the canonical quantization. An equivariant action of the inertial equivalence group can be obtained by assuming that the inertial parameter in the extended phase-space is quasi-isotropic. This assumption, combined with the general expression of the canonical diffeomorphisms, replaces the Galilei action on the space-time coordinates by the Poincar\'e transformations. \\ \indent 
These results show that the extension of the phase-space by a pair of two new canonical variables represented by energy and time is consistent with the electromagnetism and relativity. The equivariance condition indicates that the extended phase-space also provides the appropriate framework to relate the space-time geometry with quantum mechanics. The transformations of the space-time coordinates between inertial frames in relative motion appear closely related to a dynamical quantity, which is the inertial parameter in the extended phase-space. However, while space-time is classically defined, the mass of an elementary system should be defined in terms of the free dynamics of the quantum particles. The spontaneous symmetry-breaking mechanism of the quantum field theory relates the mass to the structure of the physical vacuum, and may provide a suitable basis to understand the origin of the quasi-isotropy. \\[1cm] 
 
\end{document}